\documentclass[10pt,a4paper]{article}
\usepackage{graphicx}

\usepackage{authblk}
\usepackage{fullpage,amsfonts,setspace}

\usepackage{subcaption}

\usepackage{amstext}
\usepackage{url}

\title{Urban retail dynamics: insights from percolation theory and spatial interaction modelling}

\author[1,*]{D. Piovani}
\author[1]{C.Molinero }
\author[1,2]{A. Wilson }

\affil[1]{\small{Centre for Advanced Spatial Analysis (CASA), University College London (UCL), 90 Tottenham Court Road , London, W1T 4TJ}}

\affil[2]{\small{The Alan Turing Institute, British Library, 96 Euston Road, London NW1}}

\affil[*]{d.piovani@ucl.ac.uk}

\begin{document}

\maketitle
\begin{abstract}
The study of the properties and structure of a city's road network has for many years been the focus of much work, as has the mathematical modelling of the location of its retail activity and of the emergence of clustering in retail centres. Despite these two phenomena strongly depending on one another and their fundamental importance in understanding cities, little work has been done in order to compare their evolution and their local and global properties. The contribution of this paper aims to highlight the strong relationship that retail dynamics have with the hierarchical structure of the underlying road network. We achieve this by comparing the results of the entropy maximising retail model with a percolation analysis of the road network in the city of London. We interpret the great agreement in the hierarchical spatial organisation outlined by these two approaches as new evidence of the interdependence of these two crucial dimensions of a city's life. 
\end{abstract}

\section*{Introduction}

A well known fact in the study of cities is that retail activities tend to agglomerate.  Understanding and describing this phenomenon has interested scientists from different backgrounds for many years, \cite{huff1966programmed,christaller1966central,brown1987perceptual,wilson1969use,mcfadden1980econometric}, but despite this multidisciplinary effort, in the last decades, fresh approaches have struggled to emerge. Recent advances in spatial networks \cite{Barthelemy2011,louf2013emergence}, and in road networks in particular \cite{barthelemy2008modeling,porta2006network, Cardillo2006, Louf2014}, and the large increase of available data in urban systems, have renewed the interest in the field with efforts aimed at modelling and measuring the formation of retail agglomerations \cite{ jensen2005aggregation, Jensen2006, fiasconaro2016spatio}, and relating it to centrality measurements of the city's road network \cite{Porta2009, Porta2012}.

The retail model introduced in \cite{wilson1967statistical}, a benchmark in its field, describes flows of spending power, or money,  from population centroids to retail centres. For decades this model has proven itself successful in predicting the behaviour of retail centres' dynamics \cite{Wilson1983}. 
In the model, retail centres compete for the limited amount of resources,  represented by the population, and only the more \emph{attractive} and better positioned manage to \emph{survive}. This is elegantly done through an entropy maximising model \cite{wilson2011entropy}, which quantifies the aggregate flow from population centroid $i$ to retail centre $j$ with only two parameters: one that sets the scaling between a retailer's attractiveness and floorspace and another which defines the cost of moving.  Indeed, highly visited retail centres grow proportionally to the number of visits, while poorly visited centres shrink in size and eventually are removed from the system. The identity, number and position of retail centres that \emph{survive}, strongly depends on the values of the two parameters, and it has been shown that the model undergoes a phase transition\cite{dearden2011framework,wilson2011phase,wilson2008boltzmann} from a diverse and heterogenous retail landscape to one where only the most attractive centre, by defeating all other competition, manages to survive. In between those two extreme cases the model describes the formation of retail clusters.

Moreover, in \cite{arcaute2016cities} it has been shown how the road network contains footprints of the socio-economic and cultural evolution of a country and its regions. This has been done by applying percolation theory to the network of the street intersections in the UK, which allowed to clearly uncover regional economical patterns in relation to their infrastructure.  In this approach clusters are the outcome of some thresholding process and reveal a hierarchical organisation, which is in outstanding agreement with the historical evolution of the same regions and country. 

In this paper we want to repeat the same analysis done in \cite{arcaute2016cities}, but at the city level, on London street's network, in order to study and compare the road clusters with the retail clusters that emerge from the model.
At a macroscopic level the way road clusters merge is very similar to what happens between retail clusters in the model. A low threshold scenario where many small \emph{clusters} scattered through the system appear, corresponds to the parameter values that form a heterogeneous and varied retail landscape, while the formation of the giant road cluster, corresponds to the configuration consisting of only one large retail centre.  It is therefore tempting, to bridge these two formalisms and to interpret the formation of retail clusters, described by the model, as a fingerprint of a hierarchical organisation in the economic activities of a city. As we will show in great detail the two approaches describe a very similar urban hierarchical structure, in that the spatial distribution and size of the clusters are in great agreement. This result seems to bring new evidence on the polycentric organisation of the city, and indeed sheds new light on the relationship between the road network of a city and the economical activities that develop on it. 

\section{Material and Methods}

In this section we will go through the details of the methodologies we want to compare, and the data used both for calibration and testing. We will start by defining the retail model, and analysing its main results, to then present the application of the percolation process on London’s road network. Finally, we will comment on the data sources and the calibration procedures used.

\subsection{The Retail Model}

By following the procedure outlined in \cite{wilson2008boltzmann}, we can define the flows from population centroids $i$ to retail centres $j$ as described by the equation
\begin{equation}
T_{ij}  =\frac{ p_i w^\alpha_j \text{exp}(-\beta c_{ij} )} {Z_i}
\label{eq:flow}
\end{equation}
where $p_i$ is the population in the origin $i$, $w_j$ is the aggregated floorspace of retail center $j$, and $c_{ij}$ the cost of moving from $i$ to 
$j$, we will simply quantify it as the distance. We can see how these flows are defined by two parameters, namely $\alpha$, which sets the scaling between the \emph{attractiveness} and the floorspace of a retailer and  $\beta$ which tunes the \emph{cost of moving}.  $Z_i$ is the normalisation factor, which under the constraint of the total outflow being equal to the population, i.e. $\sum_j T_{ij} = p_i$, becomes
\begin{equation}
Z_i = \sum_k w^\alpha_k \text{exp}(-\beta c_{ik})
\label{eq:Z}
\end{equation}
As one can see in \cite{wilson2008boltzmann}, the form of the flows in eq.(\ref{eq:flow}) comes out of an entropy maximising process, and is obtained under the constraints that come from the observed data: the population's $p_i$, the aggregated floorspace's $w_j$ and the cost matrix's $c_{ij}$ spatial distribution. 
This means that the set of flows $\{T_{ij}\}$ are an equilibrium configuration, that depends on the input data as well as on the values of the parameters $\alpha$ and $\beta$. Any small change in the input data would yield a rapid reconfiguration to a new attractor state. We can therefore interpret this process as a \emph{fast dynamics} one.  

Moreover we can, by exploiting eq.(\ref{eq:flow}) and eq.(\ref{eq:Z}), predict the evolution of the floorspace distribution $\{w_j\}$, considered constant during the \emph{fast dynamics}.
By calculating the total inflow to retail centre $j$ as $d_j = \sum_i T_{ij}$, we define the dynamics equation as
\begin{equation}
\Delta w_j = \epsilon (\kappa d_j - w_j) 
\label{eq:dw}
\end{equation}
which tells us that $w_j$ will increase if $\kappa d_j > w_j$ and shrink in the opposite case. The constant $\kappa$ is there to make sure that  all quantities are measured in commensurate units, and converts the flow of people into floorspace. Its value must be calibrated on the data.  

The solution to eq.(\ref{eq:dw}) is  given by the set of equations $\kappa d_j =w^\text{eq}_j$ which explicitly become
\begin{equation}
\kappa\sum_i \left\{ \frac{p_i (w^\text{eq}_j)^\alpha  \exp(-\beta c_{ij})}{\sum_k (w^\text{eq}_k)^\alpha
\text{exp}(-\beta c_{ik}) }  \right\} = w^\text{eq}_j
\label{eq:diff}
\end{equation}
The set of equations in eq.(\ref{eq:diff}) are complicated non linear equations that can only be solved iteratively. This is because every variation in any retailer's floorspace $w_k$, modifies all other equations.

\begin{figure}[t!]
\includegraphics[width=\textwidth]{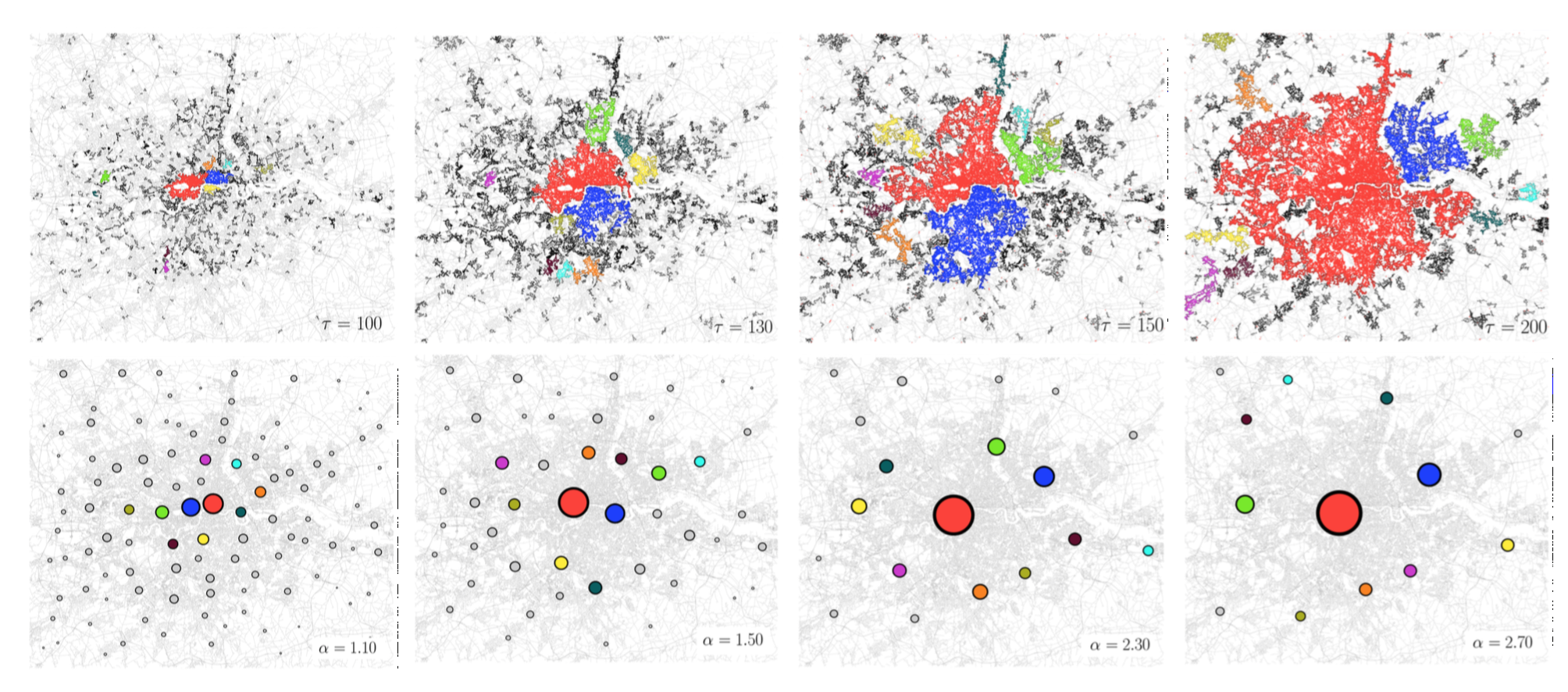}
\caption{The figures in the top panel represent the evolution of the road clusters for growing values of the threshold $\tau$. The figures in the bottom panels show the distribution of the equilibrium configuration defined in eq.(\ref{eq:dw}), $w^\text{eq}(\alpha,\beta)$, for growing $\alpha$'s and fixed $\beta = 0.8$. The size of the circles are proportional to the total floorspace of the retail cluster. In both cases the colours indicate the rank of the clusters. There is a striking similarity both in the types of dynamics, and in the spatial distribution. }
\label{fig:HierarchicalEvo}
\end{figure}
The model we have just defined has a rich behaviour and describes different types of retail structures $\{w^\text{eq}\}$, according to the two parameters $\alpha$ and $\beta$. For $\alpha > 1$ larger shops will be more attractive and a small $\beta$ implies higher probabilities of interaction over longer distances to achieve the benefits of size. Hence large $\alpha$ and small $\beta$ combinations generate structures with a small number of large $w_j$ centres and vice versa. In the bottom panels of fig.(\ref{fig:HierarchicalEvo}) we can see how, by fixing $\beta = 0.8$ and increasing $\alpha$, the number of retail centres decrease with the ones remaining becoming larger and larger.

\subsection{Percolation on London's road network}

Percolation processes \cite{Stauffer1994} are a highly studied field of research given their multiple applications to several and very distinct fields.  We can find percolation-like processes in fields that space from oil extraction \cite{King1999} to the study of the electrical conductivity of materials \cite{Clerc1990}, from polymerization processes \cite{Coniglio1979} to fire spreading \cite{Christensen1993}, from epidemiology \cite{Newman2002b}, to other health aspects such as obesity \cite{Gallos2012collective} and indeed they are used to study the structure of brain networks \cite{Gallos2012small}. 

In this work we apply a percolation process to London's road networks in order to uncover its hierarchical structure, following the procedure used in \cite{arcaute2016cities}. In the approach, the nodes of the weighted network are the road intersections, while the 
links  are the roads joining two intersections. These are weighted by their lengths, so two intersections connected by a long road will have a  link with a high weight connecting them and vice versa. The approach undertaken to calculate the percolation of London's road network consists of the following steps: we begin by setting a threshold $\tau$, then we select every link who's weight falls below that threshold, $r_{ij} < \tau$ and extract the subgraph formed by those links. The weakly connected components of the subgraph are the clusters of the network generated by the percolation process for a given threshold.  The clusters are constructed such that they have at least a link connecting them with a weight smaller than the given threshold.  These clusters form a tree structure given that for two thresholds $\tau_1$ and $\tau_2$, if $\tau_1 < \tau_2$, a cluster generated using $\tau_1$ will be completely contained into a cluster obtained using $\tau_2$. This allows us to construct a hierarchical tree that follows the ordering of the regions induced by the percolation which uncovers the intrinsic structure of the system.

Percolation is a critical process, that presents a phase transition at a critical probability (in our case threshold). This phase transition coincides with the maximum entropy of the distribution of the cluster sizes, just before the giant cluster takes over and the sizes of the secondary clusters drop to a negligable size. As we can see in fig.(2e)  the threshold that generates the maximum entropy configuration  is $\tau=130m$ and that is the point where the clusters are simultaneously maximizing their sizes while equilibrating their differences. Below it, the clusters are small, while  above it the giant cluster starts to take over the whole distribution. It is therefore, the threshold with a larger information in terms of the distribution of the cluster sizes.  The clusters of the percolation separate the network into regions that have a similar density of intersections. Considering that the population is located in buildings which are located into streets that meet at intersections, it is easy to imagine that those clusters actually correspond to concentrations of population. The study undertaken in this paper, that of finding a correspondance between the retail model and the percolation of the network structure is, therefore, one that equates the clusters of the percolation with the location of the population that in turn attracts retail centres to the centroid of their masses.

\subsection{Data and Calibration}

We have gathered data on the retailers in London from the Valuation Office Agency 2010 dataset (https://www.gov.uk/government/organisations/valuation-office-agency). There we have found information on more than $10^5$ retail activities scattered on 20707 different post-codes and for each post-code we have summed the floorspace of the retailers that belong there, and assigned it to its position. This results in $n_r =  20707$ different retail centres, which is a level of detail never tested before using the retail model in \cite{wilson1967statistical}. Furthermore we use census data at the LSOA level for the population centroids. As mentioned in the previous section, $c_{ij}$ is just a distance matrix which we were able to fill in, given the data on the spatial distribution of the population centroid and the retailer centres at the post-code level. 

Furthermore by knowing the total amount of floorspace and the total population in the city we can calibrate $\kappa$ 
\begin{equation}
\kappa P\text{tot} = F\text{tot} \longrightarrow \kappa = \frac{ F\text{tot}}{ P\text{tot}} = 1.12
\end{equation} 
where we have assumed the wealth as uniformly distributed among the population. 
  
Furthemore we have obtained the road network of London and its surroundings from the OS OpenRoads dataset \cite{OpenRoads}. This dataset is ideal, not only because it is open-source while maintaining the correct topology of the network thanks to the effort of the people at Ordnance Survey, but also because it comes pre-simplified in the sense that lanes are collapsed into centre-lines of roads. The network has been further simplified by collapsing details such as roundabouts and removing all nodes of degree 2 leaving purely the structure of the network. The final network contains 365967 nodes and 438375 links.

\section{Results}

As explained in the introduction, our aim is to compare the evolution and the spatial distribution of the road clusters emerging from the percolation process with those described by  the model we have just presented. This will allow us then to study the extent of the agreement of the hierarchical structure described by these two different formalisms. 
We will start by a qualitative observation of the two evolutions:  in Fig.(\ref{fig:HierarchicalEvo}) we show the evolution of the spatial distribution of the clusters in the two approaches. The figures in the top panel show the evolution of the percolation clusters, where nodes of the same colour belong to the same cluster, and where the colour indicates the rank of the cluster. In the bottom panels, following the same logic, we present the $\{w^\text{eq}\}$ configuration that comes out of eq.(\ref{eq:dw}), where we have fixed the $\beta$ parameter to $0.8$ and where we vary the value of $\alpha$.  At a first glance  we can see how by increasing $\tau$, which will always be measure in meters, in the road percolation approach and by fixing $\beta$ and increasing $\alpha$ in the retail model the behaviour is very similar: in both cases the number of cluster decreases while their size tends to increase, and high ranked clusters tend to position themselves in similar positions (see fig.(4a) for an overlap of the two configurations).

\begin{figure}[t!]
\includegraphics[width=\textwidth]{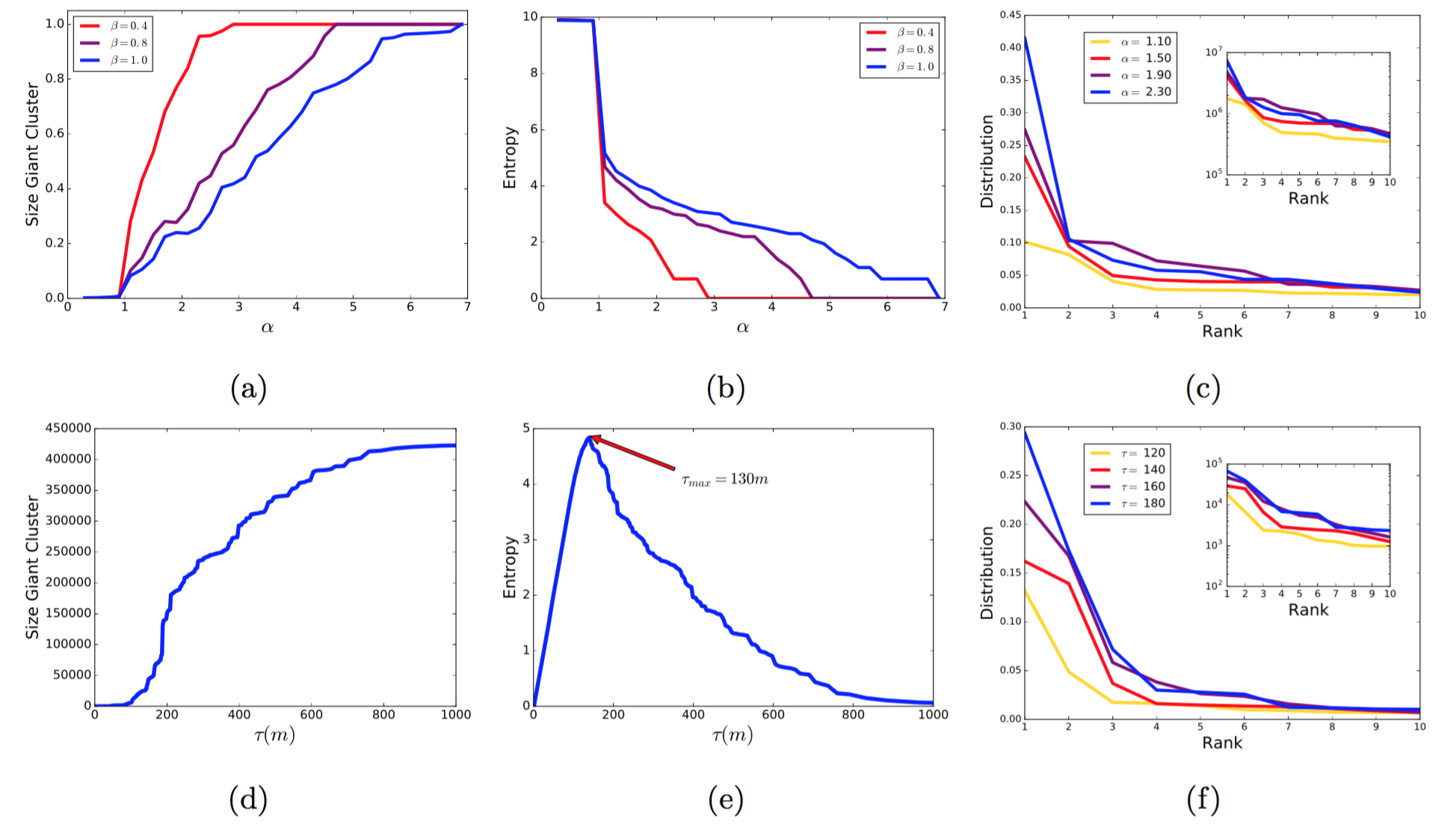}
\caption{ The top panels refer to the retail dynamics with $\beta = \{0.4 , 0.8, 1.0\}$, while the bottom panels to the hierarchical percolation. \textbf{(a)-(d)} The evolution of the size of the giant cluster, is presented, for increasing values of $\tau$ and  $\alpha$.\textbf{(b)-(e)} We show the evolution of the spatial entropy. Both quantities show a very similar behaviour in the two approaches. \textbf{(c)-(f)} We can see the distribution of the cluster that have an exponential form as shown in the insets. }
\label{fig:Perc}
\end{figure}

For a more quantitative comparison of the macroscopic properties of the two evolutions we study the size of the \emph{giant} cluster and of the entropy of the cluster sizes for increasing values of $\alpha$ and $\tau$. To calculate the entropy we have used Shannon's formula 
\begin{equation}
H_{\tau , \alpha} = - \sum_x p_{\tau,\alpha}(x) \log p_{\tau,\alpha}(x)
\end{equation}
where $x$ runs on the cluster sizes and $p(x)$ is the probability of finding a cluster of size $x$, for each value of $\tau$  and $\alpha$. In Fig.(\ref{fig:Perc}) we can see how the evolution of these two quantities follows the same behaviour in both approaches. In the percolation on the road network (bottom panels)  fig.(2d) and fig.(2e),, for low values of $\tau$, increases in the threshold imply increases in the entropy. This corresponds to a slow increase in the size of the giant cluster. Around $\tau=130$ the Entropy reaches its maximum and we can see a change in the curvature of the giant cluster which starts a steeper increase. From then on as one may expect the entropy of the system decays to zero and the giant cluster \emph{spreads} to the whole network. 

The top panels in fig.(2a) and fig.(2b), show how the clusters that form during the dynamics of the retail model follow a very similar dynamics. In this case, for all values of $\beta$, both the decrease in entropy and as well as the increase in the \emph{giant cluster} size are very slow for $\alpha < 1$.  One can see a clear transition happening at $\alpha = 1$ for any value of $\beta$ we have tested, although the type of transition depends on its specific value.  Higher values of $\beta$ yield smoother transitions, or in other words \emph{converge} for higher values of $\beta$, while for lower values of $\beta$ we get sharper transitions.  However no matter the value of $\beta$ the system always ends up with the same winner: the same retail cluster manages to outplay the rest of the competitors and have all the flows in the systems directed towards it. This means that for each value of the parameter we are always observing the same transition, which begins from the same initial condition and ends up in the same \emph{ground state}. We could think of $\beta$ as setting the scale of the transition, in terms of $\alpha$: the system always \emph{explores}  the same states, but  in a low $\beta$ scenario one needs more coarse grained values of $\alpha$ than in a high $\beta$ system to actually observe them all. This allows us to fix the value of $\beta$ (we will arbitrarily set to $\beta = 0.8$) and study the behaviour of the system only varying the values of $\alpha$. In the two figures on the right fig.(2c) and fig.(2f) we show the distribution of the sizes of the clusters, i.e. the aggregated floorspace for the retail centres and number of nodes on the road network clusters, of the 10 largest clusters. In the insets we can see how the distribution has an exponential form in both cases, and how increasing $\alpha$ and $\tau$ has the same effect on the distribution, namely increasing its steepness. 

\begin{figure}
\centering
\includegraphics[width=\textwidth]{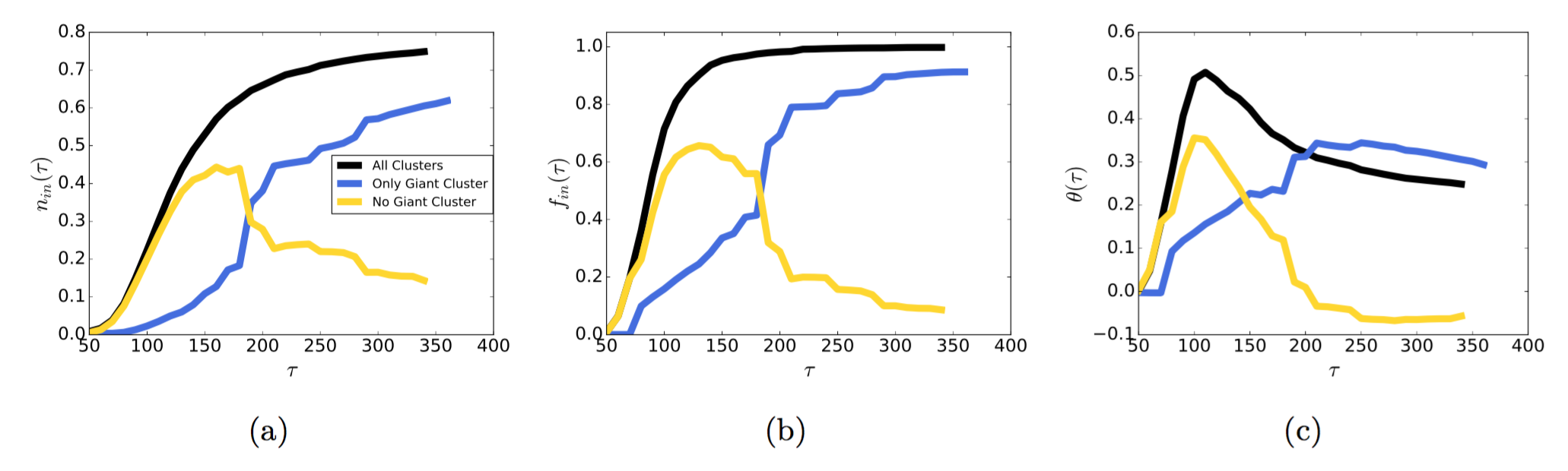}
\caption{\textbf{(a)} This figure represents the evolution of the fraction of nodes of the road network allowed in the system for the given value of $\tau$, $n_\text{in}(\tau)$ (black curve). Moreover we show the same quantity considering only the giant cluster (blue curve) and every cluster but the giant cluster (yellow curve). \textbf{(b)} Here we show the fraction of the total retail floorspace belonging to the nodes that form the clusters $f_\text{in}(\tau)$ where the colours have the same meaning. \textbf{(c)} We compare the two curves by showing their difference $\Theta(\tau)= (n_\text{in} - f_\text{in})$}
\label{fig:Full_Ret}
\end{figure}

In this paragraph we have seen how at a macroscopic level the two approaches describe a very similar dynamics in the formation of clusters, and eventually of a \emph{giant cluster}. We have also shown how we can fix the $\beta$ without loss of generality in the results and how, in the retail model,  the $\alpha$ parameter plays the same role as the threshold $\tau$ plays in the percolation. In the following we will measure the spatial similarity of the cluster's distribution. 

\subsection{Retail distribution on percolation road clusters}

Before moving to a more detailed comparison and analyse their local distribution, we must take a step back, and see how the retail centres we found in the data are distributed on the percolation road clusters bearing in mind that it represents the initial condition for the model's evolution. This step is interesting for two reasons: on one hand it will tell us if we can learn something on the real retail distribution by analysing its relationship to the road clusters, and on the other hand it will serve as a benchmark to then better quantify the effects of the retail model's dynamics.  In \cite{arcaute2016cities} the authors have shown how  starting from the road network of the whole of the UK, the cities emerged as  clusters of the road network for  $\tau = 300m$. Given that our analysis is applied at the city level in London, we will take that as our maximum threshold. Furthermore we have considered as the minimum size of a cluster to be allowed in the system as $n_\text{road} = 50$.

We consider the retail floorspace spatial distribution aggregated the post code level, and assign it to the closest node in the road network, and for each
 $\tau$ we then study the fraction of floorspace assigned to the emerging clusters. To do so we compare the fraction of nodes of the road networks in the system of a given threshold, namely 
 \begin{equation}
n_\text{in} (\tau)= \frac{n_\text{c}(\tau)}{N_\text{road}} 
\end{equation}
where $n_\text{c}(\tau)$ is the sum of nodes that form all clusters in the system and $N_\text{road}$ the total number of nodes to the amount of retail floorspace \emph{contained} in them:
\begin{equation}
f_\text{in} (\tau) = \frac{f_\text{c}(\tau)}{F_\text{tot}}
\end{equation}
where once again $f_c(\tau)$ is the amount of floorspace assigned to the nodes that belong to the percolation's clusters and $F_\text{tot}$ the total amount of floorspace in the system. We can then study their difference 
 \begin{equation}
 \Theta(\tau) =( n_\text{in}  - f_\text{in})
 \end{equation}
A $\Theta \simeq 0$ case would indicate a random spatial distribution of the retailers on the road network. Meaning that the distribution of retail floorspace on the road network would be independent of the hierarchy indicated by  $\tau$, and one would obtain the same $f_\text{in}$ by selecting the same fraction of nodes, $f_\text{in}$, using any other criteria.  On the other hand $\Theta < 0$, would indicate a tendency of retailers to be on roads that are not yet in the system, while the opposite case $\Theta >0 $ would unveil a spontaneous tendency of retailers in positioning themselves on highly connected clusters.

In Fig.(\ref{fig:Full_Ret}) we show the behaviour of the distribution of the three quantities $n_\text{in}(\tau)$, $f_\text{in}(\tau)$ and $\Theta(\tau)$ in fig.(3a), (3b), (3c) respectively. We have measured the quantities on the full network (black curve), the network without the giant cluster (yellow curve) and only considering the giant cluster (blue curve). This has been done to make sure the results where not being dominated by the giant cluster. By comparing the figures we can see how $f_\text{in}$ grows much faster in $\tau$ with respect  to $n_\text{in}$, and we constantly get $\Theta > 0$. Furthermore the red curve shows that up to $\tau=200$, this is true even if we exclude the giant cluster from the analysis. If it is clear from these results that retailers tend to position themselves in central locations, what also emerges is a tendency to choose highly connected clusters, or in other words clusters  formed by a dense grid of alleys and road intersections. An in depth analysis of these results would require a study of its own and we leave it to future research, and will now exploit these results to understand the effects introduced by the dynamics described in eq.(\ref{eq:flow})-(\ref{eq:diff}). 
\begin{figure}[]
\centering
\begin{subfigure}[b]{\textwidth}
\centering
        \includegraphics[width=.8\textwidth]{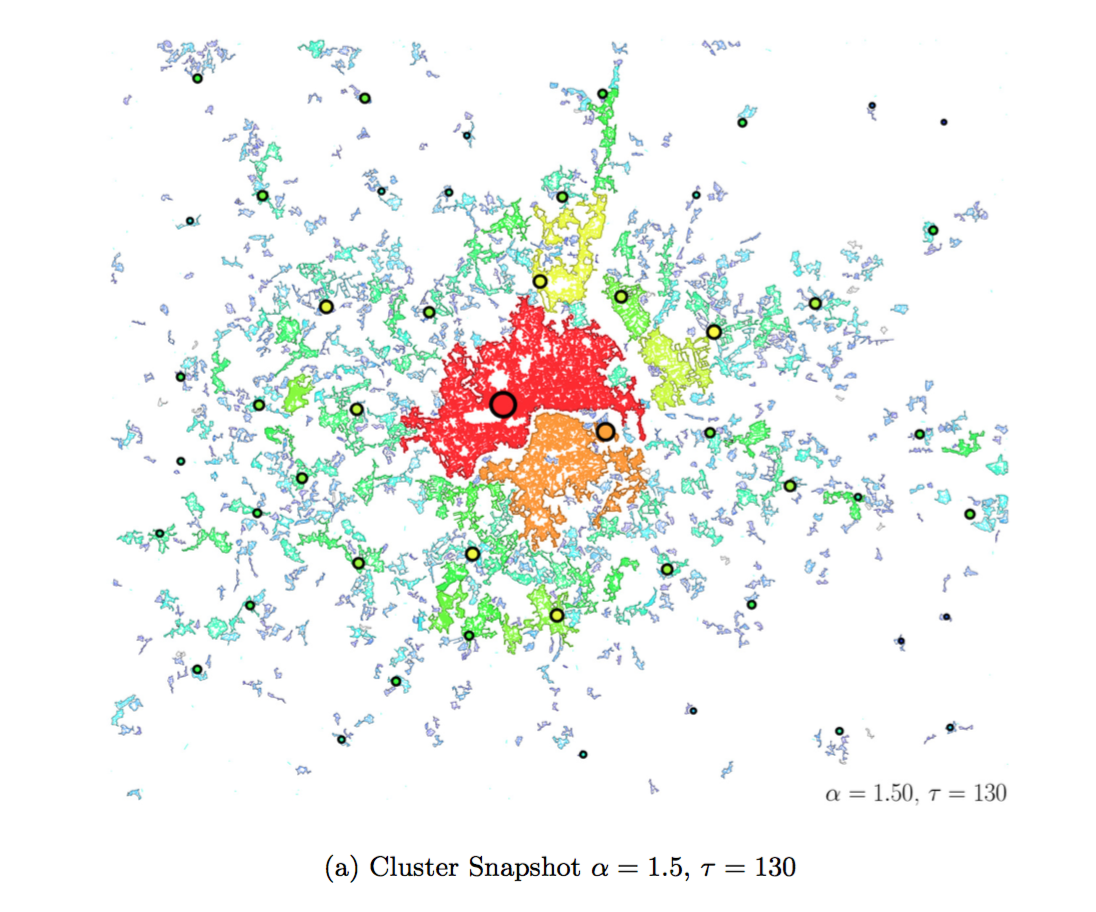}
        \label{fig:snapshot}
\end{subfigure}
\begin{subfigure}[b]{\textwidth}
	\centering
        \includegraphics[width=0.8\textwidth]{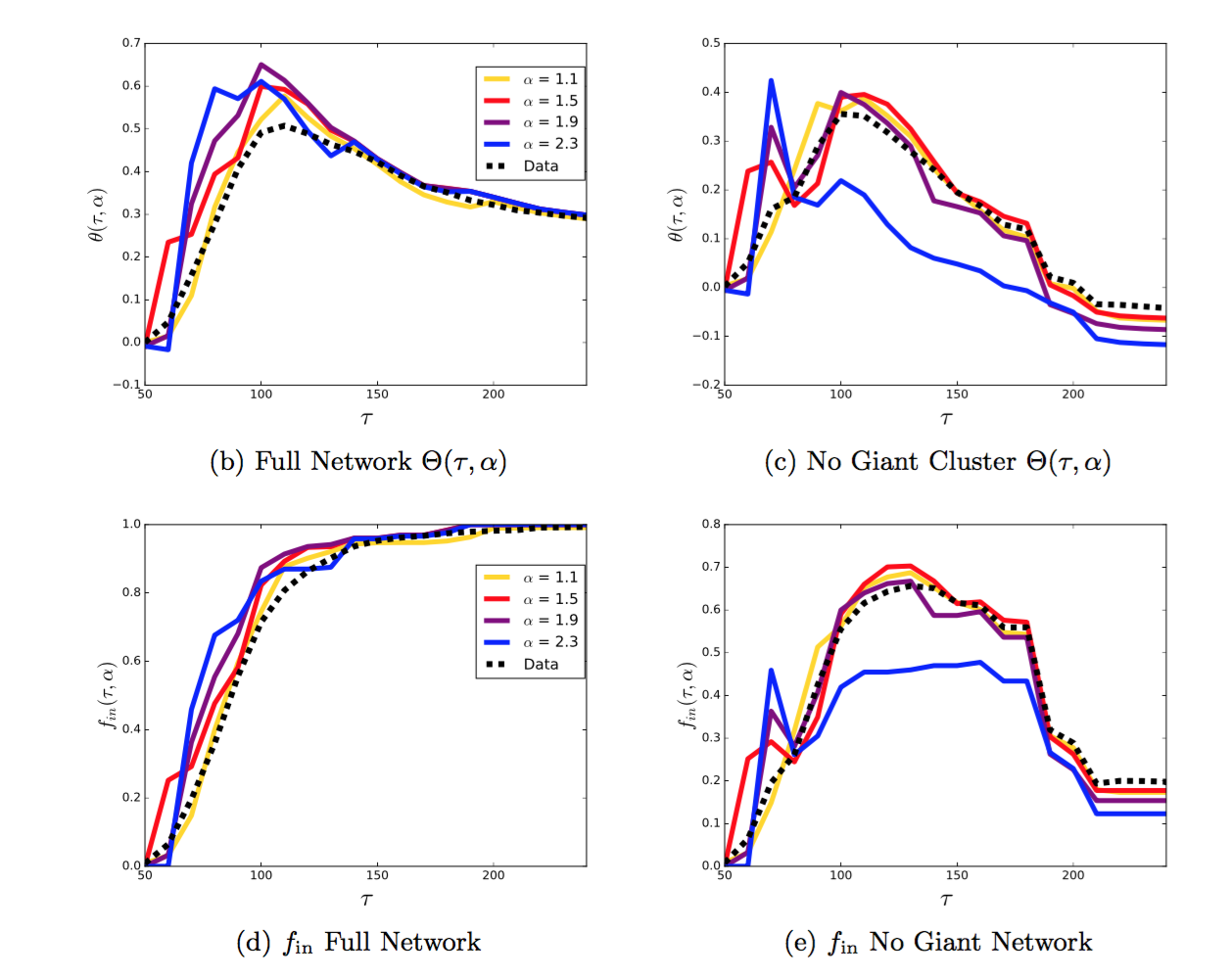}
        \label{fig:ta}
\end{subfigure}
\caption{\textbf{(a)} We have overlapped a snapshot of the road network clusters for $\tau = 130$ with the retail clusters $\{w^\text{eq}\}$ for $\alpha=1.5$. The colours indicate the rank of the size of the clusters. We can see how most of retail clusters fall on road clusters, and there is good agreement between the spatial distribution of the ranks. In \textbf{(b)-(c)} We show how $\Theta(\alpha,\tau)$ varies with $\tau$ for several values of $\alpha$, on the full network and not considering the giant cluster. The dashed black line indicates the values obtained from the data. We can see that the model till $\alpha = 1.9$ constantly produces higher values, with and without the giant cluster, while for $\alpha=2.3$ we can see how the giant cluster plays a fundamental role. This is because for that value of $\alpha$ the floorspace is mostly concentrated in the giant cluster (see Fig.(\ref{fig:GiantRet}).\textbf{(d)-(e)} We show the $f_\text{in}(\alpha,\tau)$ with and without the giant cluster. The results are in line with that said for the previous figures. }
\label{fig:theta_alpha}
\end{figure}

\subsection{Comparing the spatial distribution of retail and road clusters.}

We have seen how the retailers are more likely to be found on roads that belong to very connected clusters, but we still did not apply the model's dynamics to the retailer's spatial distribution. We have also seen that just like in the road network percolation $\tau$ shows the hierarchical relationships of the roads, $\alpha$ in the model describes the formation of retail clusters described in $\{w^\text{eq}\}$. Now we want to study the analogies and differences of these emerging structures. We do this by repeating the same analysis we have done on the full data set, this time on the equilibrium configurations that comes out of eq.(\ref{eq:dw}). Of course, the introduction of $\alpha$ adds a new degree of freedom, and now $f_\text{in}\equiv f_\text{in}(\alpha,\tau)$ and  $\Theta \equiv \Theta(\alpha,\tau)$.

In Fig.(\ref{fig:snapshot}) we start by showing an overlap of the clusters obtained with $\beta = 0.8, \alpha=1.5$ and $\tau = 130$. At a first glance we can see how big retail clusters tend to lay on big road clusters, and vice versa, and how this is true  even for clusters appearing at the periphery of the city. Some clusters that did not exactly overlap lie one next to the other.  To quantify this impression, in Fig.(\ref{fig:tas}) and Fig.(\ref{fig:ratio_ta}) we show $ f_\text{in}(\alpha,\tau)$ and $\Theta (\alpha, \tau)$ for values of $\alpha$ ranging from $1.1$ to $2.3$. For $\alpha$ greater than that, the floorspace is mainly contained in the giant cluster which dominates any analysis. Perhaps surprisingly $ f_\text{in}(\alpha, \tau) > f_{in}(\tau) $ and  $\Theta (\alpha, \tau) >  \Theta (\tau)$ $ \forall  \tau$, indicating that retailers belonging to the clusters have \emph{survived} the dynamics more than those not belonging to the clusters, and have grown in size.  This is true both if we include the giant cluster and if we leave it out. For $\alpha=2.3$ however $40\%$ of the retail floorspace is concentrated in the retail giant cluster which lies on the road's giant cluster, therefore not considering it ruins the results.  Finally, in Fig.(\ref{fig:size_corr}) we show the correlation between the log of the amount of floorspace on a road cluster and its size. The high levels of correlation  imply that big retail clusters tend to position themselves on big road clusters, and this effect is improved by the model.
\begin{figure}[]
\centering
 \includegraphics[width=\textwidth]{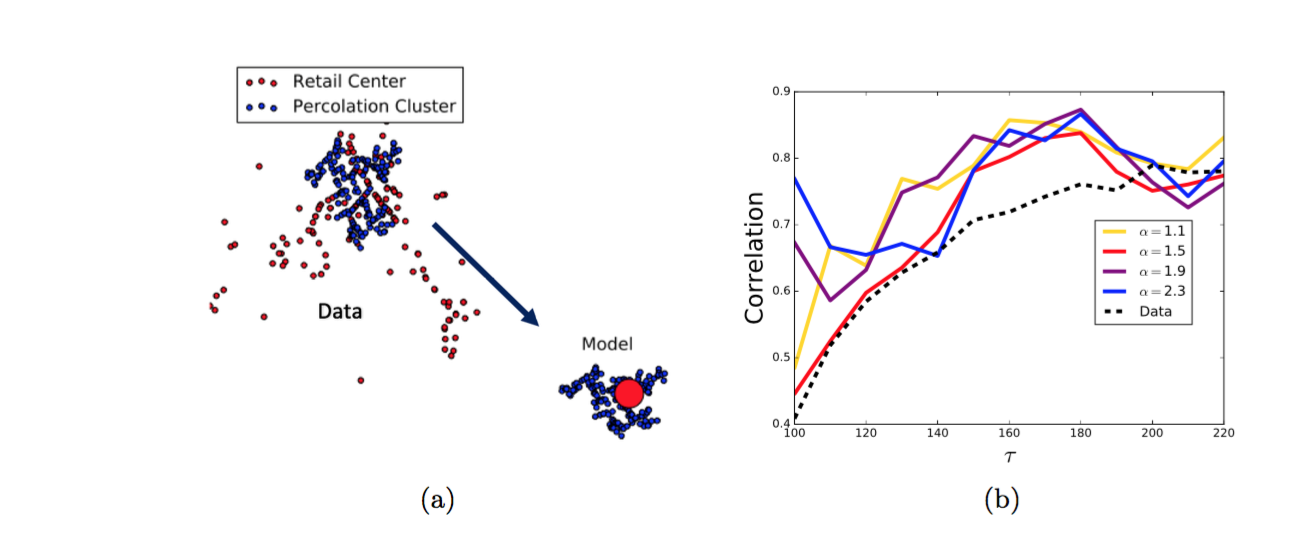}
 \caption{\textbf{(a)} A diagram of the effect of the model's dynamics on the retail floorspace distribution. We see how the retail floorspace distribution in the data falls outside the road cluster, and how after the dynamics, the floorspace is all concentrated inside the cluster. \textbf{(b)} The effect shown in (a) can be seen by the higher levels of correlation between the sizes of the road cluster and the amount of floorspace belonging to it. }
\end{figure}\section{Discussion and Conclusion}

Cities have been successfully characterised both by analysing the structure of their road networks \cite{porta2006network,Louf2014,Murcio2015,barthelemy2008modeling} and the distribution and nature of their retail activity \cite{wilson1967statistical,mcfadden2001economic, um2009scaling}. Some work has been done to relate these two approaches \cite{Porta2009,Porta2012} and the contribution presented in this paper goes in that direction.  The presence of a hierarchical structure in its road networks has been showed \cite{arcaute2016cities} as has the mechanism that leads to it \cite{louf2013emergence}. Percolation theory has proven itself as a useful tool to study urban areas \cite{arcaute2016cities,Li2015} but to our knowledge no work has been done to export these concepts to analyse the organisation of retail activities in cities.  To do so we have used the single constrained entropy maximising retail model \cite{wilson1967statistical}, perhaps the most widely used model of the field, and characterised its results by \emph{embedding} them on the city's  road network.  We have used the city of London as a test case, and given the quality of our results now plan to extend our research to the whole of the UK. 

In other words we have presented an attempt to relate the different configurations of clusters obtained through purely geometrical means based on the road network with the evolution described by the retail model. We have quantified their  agreement by measuring the amount of retail floorspace contained within the clusters and studied how far this is from a random distribution. Furthermore we have studied the correlation between the size of the clusters and the amount of retail floorspace they contained.  As we have seen  the configurations obtained by using these two formalisms are very similar, with the spatial distribution of retail clusters being very close to the distribution of road clusters, and the correlations levels we have found are very high. By comparing these results with those obtained using  the spatial distribution of retailers in London aggregated at the post code level, we have shown how the model improves the agreements, which therefore does not depend on the original distribution. We believe that the results presented in this paper are important for a number or reasons. We have bridged the results of a model first presented many years ago with an approach that only recently has been applied to study urban spaces, and comparing the results we have interpreted them in a new way. Indeed we have brought evidence to the existence of a  hierarchical spatial organisation of retail activity in London, and believe this result can be very useful in future modelling.

\section{Acknowledgements}
The authors wish to thank Elsa Arcaute and Michael Batty for insights and comments, and Stanislao Gualdi for precious suggestions on the presentation of the results.The authors wish to acknowledge the support of the EPSRC grant: EP/M023583/1.

\bibliographystyle{unsrt}
\bibliography{percolation_bib}

\end{document}